\documentclass[journal=jctcce,manuscript=article,layout=twocolumn]{achemso}

\usepackage{amsmath}
\usepackage{xcolor}
\usepackage{braket}
\usepackage{array}
\usepackage{graphicx}
\usepackage{subfig}
\usepackage{caption}
\usepackage[singlelinecheck=false]{caption}

\author{Pavlo Golub}
\email{pavlo.golub@jh-inst.cas.cz}
\affiliation[jh-inst]
{J. Heyrovsky Institute of Physical Chemistry, v.v.i., Czech Academy of Sciences, Prague, Czech Republic}
\author{Chao Yang}
\affiliation{Applied Mathematics and Computational Research Division, Lawerence Berkeley National Laboratory, Berkeley, USA, 94720.
}
\author{Vojt\v{e}ch Vl\v{c}ek}
\affiliation{Department of Chemistry and Biochemistry, University of California,
Santa Barbara, Santa Barbara, USA, 93117 and
Department of Materials, University of California, Santa Barbara, Santa Barbara, USA, 93117.
}
\author{Libor Veis}
\affiliation[jh-inst]
{J. Heyrovsk\'y Institute of Physical Chemistry, v.v.i., Czech Academy of Sciences, Prague, Czech Republic.}
\email{libor.veis@jh-inst.cas.cz}

\title[An \textsf{achemso} demo]
  {Quantum Chemical Density Matrix Renormalization Group Method Boosted by Machine Learning} 

\begin{document}

%\begin{tocentry}

%\end{tocentry}

Accurate electronic structure calculations are essential in modern materials science, but strongly correlated systems pose a significant challenge due to their computational cost. Traditional methods, such as complete active space self-consistent field (CASSCF), scale exponentially with system size, while alternative methods like the density matrix renormalization group (DMRG) scale more favorably, yet remain limited for large systems. In this work, we demonstrate how a simple machine learning model can enhance quantum chemical DMRG calculations, improving their accuracy to chemical precision, even for systems that would otherwise require considerably higher computational resources. The systems under study are polycyclic aromatic hydrocarbons, which are typical candidates for DMRG calculations and are highly relevant for advanced technological applications.

The concept of using machine learning (ML) to refine low-level theoretical calculations, bringing them closer to high-level accuracy, is a promising and actively evolving approach known as $\Delta$-ML \cite{Ramakrishnan-Dral-2015}. Various computational starting points have been explored in this context, including density functional theory (DFT) \cite{Smith-Nebgen-2019,Nandi-Qu-2021,Bowman-Qu-2022,Ruth-Gerbig-2023,Maier-Collins-2023,Collins-Raghavachari-2023} or Hartree-Fock (HF) single-reference calculations \cite{Cheng-Welborn-2019}, and post-HF \textit{ab initio} methods like second order M{\o}ller-Plesset perturbation thoery (MP2) \cite{Townsend-Vogiatzis-2020,Kaser-Boittier-2021} or coupled clusters with singles and doubles (CCSD) \cite{Ruth-Gerbig-2022}. Additionally, variational two-electron reduced-density matrix (v2RDM) descriptions have been used as starting points \cite{Jones-Li-2023}. These methods are optimized against highly accurate, yet computationally intensive, benchmarks such as coupled clusters with perturbative triples [CCSD(T)] or complete active space configuration interaction (CASCI), aiming to achieve results that closely approximate high-level accuracy.

The DMRG method \cite{White-1992,White-1993} is a powerful variational approach widely used for studying strongly correlated quantum systems \cite{schollwock_2005}. By optimizing the many-body wave function within a truncated Hilbert space, DMRG achieves high computational efficiency without compromising accuracy. In quantum chemistry applications \cite{White1999, chan_review, wouters_review, Szalay2015, reiher_perspective}, DMRG typically approximates the ground state (or low-lying excited states) of a full configuration interaction (FCI) solution within a chosen orbital space, such as that defined by the CASCI framework. An example can be the $\pi$-orbital active space of polycyclic aromatic hydrocarbons (PAHs) presented below. 

The DMRG algorithm provides the wave function in a matrix product state (MPS) representation, which allows for an efficient and compact description of entangled quantum states \cite{Schollwock2011}. The FCI wave function, in the occupation basis representation, is expressed as
\begin{equation}
  | \Psi_{\text{FCI}} \rangle = \sum_{\{\alpha\}} c^{\alpha_1 \alpha_2 \ldots \alpha_n} | \alpha_1 \alpha_2 \cdots \alpha_n \rangle,
\end{equation}
where $\alpha_i$ represents the occupation state of the $i$-th orbital, with $\alpha_i \in { | 0 \rangle, | \downarrow \rangle, | \uparrow \rangle, | \downarrow \uparrow \rangle }$. By successively applying SVD to the FCI tensor $c^{\alpha_1 \alpha_2 \ldots \alpha_n}$, the wave function can be factorized into an MPS form \cite{Schollwock2011}
\begin{equation}
  \label{mps_factorization}
  c^{\alpha_1 \ldots \alpha_n} = \sum_{i_1 \ldots i_{n-1}} A[1]_{i_1}^{\alpha_1} A[2]_{i_1 i_2}^{\alpha_2} A[3]_{i_2 i_3}^{\alpha_3} \cdots A[n]_{i_{n-1}}^{\alpha_n},
\end{equation}

\noindent
where $A[j]^{\alpha_j}$ are the MPS matrices corresponding to each orbital. The new indices, $i_j$, introduced by SVD, are called virtual indices, and they are contracted across different MPS matrices. If the MPS factorization was exact, the dimensions of these matrices would grow exponentially with a system size, similar to the growth of the original FCI tensor. In the DMRG algorithm, however, the dimensions of the virtual indices are truncated, resulting in the reduced computational complexity. They are called bond dimensions and are typically denoted by $M$. The choice of $M$ controls the accuracy of the approximation, with larger bond dimensions capturing more entanglement at the cost of higher computational demands.

The iterative protocol of the practical two-site DMRG algorithm assumes that the orbitals are arranged in a 1D chain. The system is divided into two large blocks (left and right) with two smaller blocks, each consisting of a single orbital, positioned between them. The algorithm performs a sweeping process from left to right, gradually enlarging the left block by one orbital while shrinking the right block by the same amount. Once the end of the chain is reached, the sweep reverses direction. In each iteration of the sweep, the eigenvalue problem corresponding to the projected Schrödinger equation onto the tensor product space of the aforementioned four blocks is solved.

In the original DMRG formulation \cite{White-1992, White-1993, schollwock_2005}, the explicit determinant representations of the complex many-particle bases are not stored. Instead, the matrix representations of second-quantized operators required for applying the Hamiltonian to a (trial) wave function are constructed and retained. Transitioning between iterations during a DMRG sweep occurs via a renormalization procedure. A key component of the DMRG algorithm is truncation, achieved through singular value decomposition (SVD) of the wave function in its bipartite form, expanded in the basis of the enlarged left and right blocks

\begin{equation}
  \ket{\Psi} = \sum_{L, R} \psi_{LR} \ket{L} \otimes \ket{R}.
\end{equation}

\noindent
$\ket{L}$ and $\ket{R}$ represent the basis states of the merged left and right blocks, each including a neighboring single orbital. The $4M \times 4M$ matrix $\psi_{LR}$ is approximated  by $M \times M$ matrix $\tilde{\psi}_{LR}$ by using only the largest singluar values. Alternatively, this truncation can be achieved by diagonalizing the density matrix of the enlarged left or right block and preserving only the 
$M$ largest eigenvalues. A key indicator of the accuracy of the approximation at a particular iteration of the DMRG sweep is the truncation error (TRE), given by

\begin{equation}
\label{trunc_error}
\vert \vert \vert \Psi \rangle - \vert \tilde{\Psi} \rangle \vert \vert ^{2} = 1 - \sum_{\alpha_{L}}^{M} \langle \alpha_{L} \vert \hat{\rho}_{L} \vert \alpha_{L} \rangle,
\end{equation}

\noindent
where $\hat{\rho}_{L}$ represents the reduced density operator of the enlarged block and $\ket{\alpha_{L}}$ denote its eigenvectors. 

An important characteristic of the quantum system under study, easily accessible through the DMRG calculation, is its entanglement properties, which can be quantified using the 
$N$-orbital entanglement entropy as defined by the von Neumann entropy. \cite{Legeza-Solyom-2003,Legeza-Solyom-2004,Rissler-Noack-2006}.

\begin{equation} \label{1p_entrop}
    s^{(N)} = -\sum_{\sigma}{w_{\sigma;1...N} ~ \textrm{ln} (w_{\sigma;1...N})},
\end{equation}

\noindent
where $w_{\sigma;1,...N}$ represent the eigenvalues of the reduced $N$-orbital density matrix. For example a single-orbital entropy, $s^{(1)}$, quantifies the entanglement between a single orbital and the remaining subset of orbitals, while the two-orbital entropy, $s^{(2)}$, measures the entanglement between a pair of orbitals and the rest of the system. The mutual information, which reflects the correlation between a specific pair of orbitals, is given by the following expression

\begin{equation} \label{mut_info}
    I_{ij} = s^{(1)}_{i} + s^{(1)}_{j} - s^{(2)}_{ij}.
\end{equation}

%% This I would probably skip...
%Since DMRG algorithm steadily transforms from specialized to routine method as a number code implementation are already available or will be in the nearest future\cite{Brabec-Brandejs-2020,Zhai-Larsson-2023,Weymuth-Unsleber-2024}, auxiliary supporting protocols using the data available directly after DMRG calculations are desirable.

As mentioned above, TRE quantifies the accuracy of the wave function. It has been observed empirically \cite{Legeza-Fath-1996,Chan-Head-Gordon-2002,Olivares-Amaya-Hu-2015} that the maximum TRE from the last sweep before the convergence is achieved (also denoted as discarded weight) is almost linearly proportional to the error in the DMRG energy. This fact allows for extrapolations to the truncation error zero limit \cite{Olivares-Amaya-Hu-2015}. Accurate extrapolations, however, require multiple calculations with increasing bond dimensions spanning several orders of magnitude in TRE. Given the scaling of the DMRG algorithm with bond dimension, $\mathcal{O}(M^3)$, it is evident that this process can become computationally expensive.

\begin{figure*}[!ht]
  \includegraphics[width=18cm]{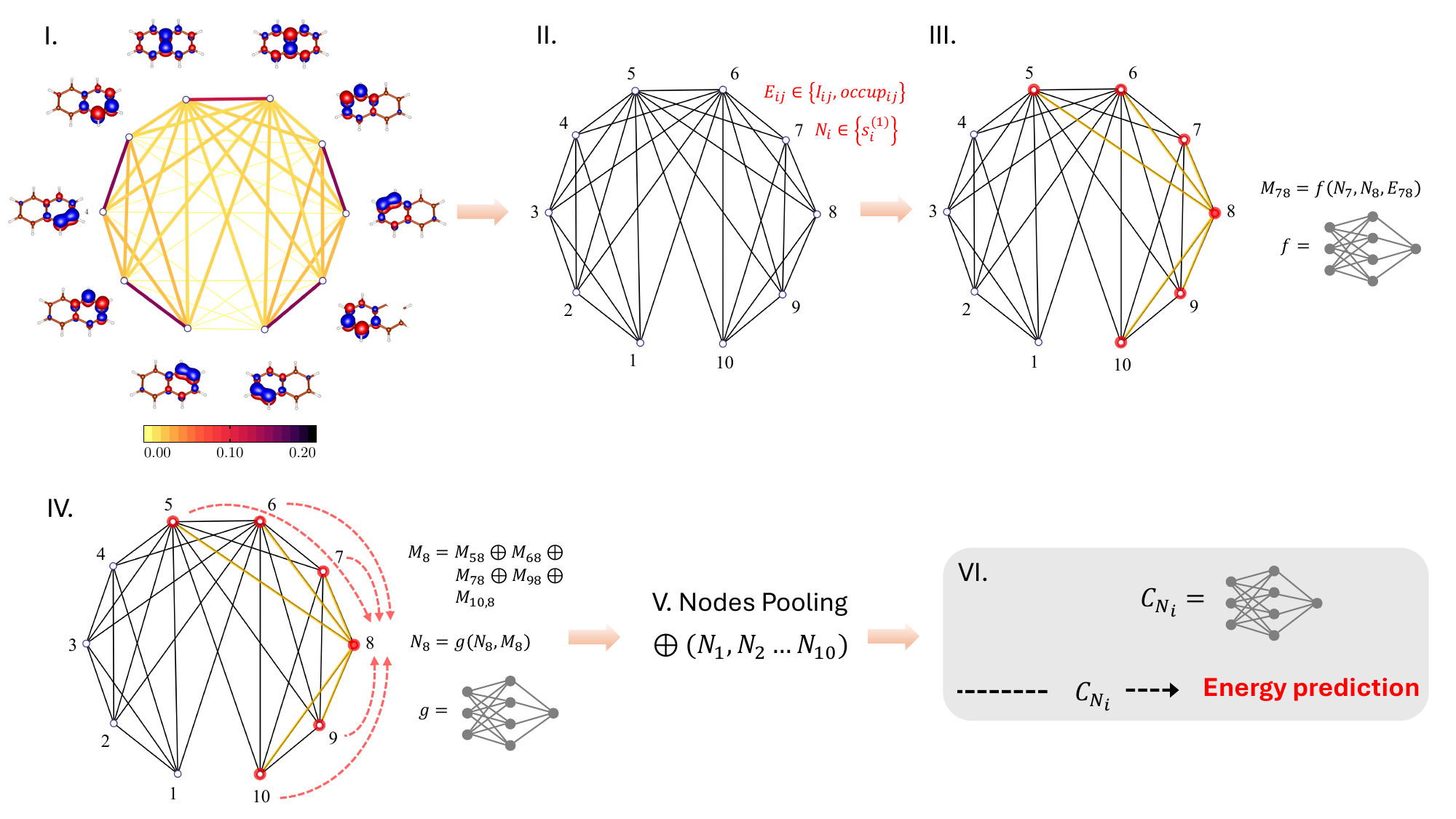}
  \caption{Schematic representation of the MPGNN model used in this work. (I) Mutual information plot. (II) Corresponding graph representation, showing node and edge features. (III) Message formation comprising neural network processing ($f$). (IV) Message update comprising message pooling and neural network processing ($g$). Steps III and IV are repeated $k$-times for a $k$-layer MPGNN. (V) Final pooling of node representations. (VI) Neural network-based energy prediction. For further details, refer to the main text.}
  \label{gnn_scheme}
\end{figure*}

% Here comes our mission!
In this letter, we introduce an alternative approach. We assume that single-orbital entropies and mutual information capture sufficient information to predict the behavior of the DMRG energy error as the bond dimension increases. In the spirit of $\Delta$-ML methods, we propose using correlation measures from calculations with significantly lower bond dimensions to estimate energies in the zero truncation error limit. Since the dependence patterns may vary across different types of correlated systems—and are often complex, non-obvious, or non-uniform—machine learning techniques are particularly well-suited to uncover and model these hidden relationships.

A way to unify the analysis of molecules of varying sizes is by representing them as graph data structures. This involves organizing the available information so that part of the data corresponds to relatively separable entities (nodes), while the remaining data is associated with pairs of nodes (edges). For instance, a natural way to represent real-space molecular structures as graphs is by treating constituent atoms as nodes and atomic bonds as edges. In machine learning, the processing of graph-structured data falls under the domain of graph neural networks (GNN) \cite{Scarselli-Gori-2009, SanchezLengeling2021}.

In quantum chemistry, DMRG is typically applied to a set of molecular orbitals. In this context, a natural way to represent the system as a graph is to treat each individual orbital as a node. Consequently, single-site entropies become node features, while mutual information (or two-site entropies) can serve as edge features. In this work, we used the mutual information value as the edge feature and set a minimal threshold of 0.004 for edge existence. Additionally, we incorporated the DFT orbital occupancy information of a pair of orbitals as another edge feature (see Supplementary Information, SI, for more details).

In our proof-of-concept study, we employ a simple message-passing graph neural network (MPGNN) approach, as illustrated in Figure \ref{gnn_scheme}.
The graph representation is constructed based on mutual information, using a predefined threshold (Step II in Figure \ref{gnn_scheme}). In the next stage (Step III), messages are formed by concatenating the node and edge features of connected neighbors. These messages are then processed by a differentiable function $f$, such as a deep neural network.
Step IV in Figure \ref{gnn_scheme} illustrates the message update process, which involves aggregating messages from connected neighbors, concatenating the aggregated messages with the node features, and processing them with another differentiable function $g$. The aggregation operation, denoted by $\bigoplus$, can use various pooling techniques, such as summation, mean, or others; in this study, we use mean pooling.

After $k$-layer message passing, the graph-level representation is constructed by applying an aggregation operator across all nodes. This graph representation is then augmented with the corresponding truncation errors and fed into a fully connected neural network, $C$, which outputs the energy prediction.
The optimization process minimized the difference between the predicted energies and the reference high-bond-dimension DMRG energies corresponding to the TRE zero limit.

% we probably do not need to be so specific...

%The optimization process minimized the difference between the low-bond-dimension ($E_{lb}$) and high-bond-dimension ($E_{hb}$) correlation energies. The target is expressed as a percentage of the energy difference between the reference Hartree-Fock energy ($E_{\text{HF}}$) and the low-bond-dimension correlation energy

%\begin{equation}
%t = 100 ~ \frac {E_{lb}-E_{hb}} {E_{\text{HF}}-E_{lb}}.
%\end{equation}

In this study, we utilized the simplest form of an MPGNN, consisting of a single message-passing layer, with $f$ and $g$ represented as identity functions. This basic MPGNN configuration is the primary focus throughout the main text. Results for more complex MPGNN architectures, along with a detailed description of both network configurations, are provided in SI.

The training dataset consisted of elements from the publicly available database of polycyclic aromatic hydrocarbons (PAHs) COMPAS-1D\cite{Wahab-Pfuderer-2022}. It included 100 molecules: all PAHs with 5 and 6 benzene rings (49 molecules in total) and selected 51 PAHs with 7 benzene rings and the smallest HOMO-LUMO gaps.

\begin{figure}[!t]
\includegraphics[width=8cm]{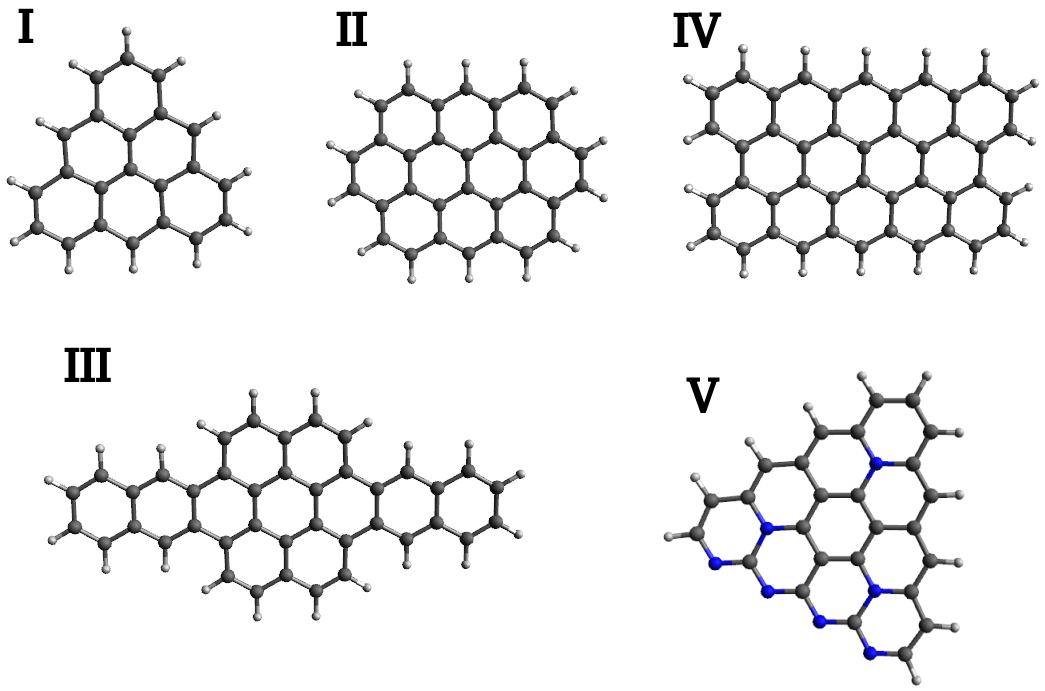}
\caption{Test set of polycyclic aromatic hydrocarbons: {\bf{I}} -- C$_{22}$H$_{12}$, [3]triangulene; {\bf{II}} -- C$_{32}$H$_{14}$, ovalene; {\bf{III}} -- C$_{40}$H$_{20}$; {\bf{IV}} -- C$_{44}$H$_{18}$, peripentacene; {\bf{V}} -- C$_{26}$N$_{7}$H$_{11}$.}
\label{fig:test_mols}
\centering
\end{figure}

The test dataset was intentionally designed to include peri-fused PAHs that are absent from the COMPAS-1D database. The selected molecules are shown in Figure \ref{fig:test_mols}. It is well known that the electronic structure of PAHs depends heavily on molecular topology \cite{Ortiz-Boto-2019, Song2024}. In order to demonstrate that our ML model is agnostic to this property, 
we also included [3]triangulene (system I), which unlike molecules in the training dataset exhibits sub-lattice imbalance, resulting in a triplet ground state. 
To evaluate performance in the presence of hetero-atoms, the dataset included the aza-analogue of [4]triangulene (system V). Furthermore, to test the model's transferability to more extended systems, we included peripentacene (system IV), the largest molecule in the test dataset (14 benzene rings), which is considerably larger then molecules in the training dataset. In addition, peripentacene exhibits the open-shell singlet ground state and extrapolated DMRG reference singlet-triplet (S-T) energy gap is available \cite{Grande-Urgel-2020}.

The geometries of the molecules in the training dataset were obtained from the COMPAS-1D database \cite{Wahab-Pfuderer-2022}, while those in the test dataset were adopted from other studies and are provided in SI. Input orbitals for DMRG calculations were computed at the DFT level using the B3LYP exchange-correlation functional \cite{Parr88_785,Becke88_3098} and the cc-PVDZ basis set \cite{Dunning1989}. 

It is well established that DMRG achieves optimal performance in a local basis \cite{Olivares-Amaya-Hu-2015}, where the MPS parametrization effectively leverages the locality of electron correlation. To ensure this, the initial DFT orbitals were split-localized using the Pipek-Mezey procedure \cite{Pipek1989}. All calculations for PAHs presented in this study employed complete $\pi$-active spaces, corresponding to CAS sizes of (22e, 22o), (26e, 26o), and (30e, 30o) for molecules containing 5, 6, and 7 benzene rings, respectively. For each molecule, three DMRG calculations were performed at low bond dimensions, resulting in truncation errors ranging from 1$\times$10$^{-4}$ to 5$\times$10$^{-6}$. High-accuracy reference calculations achieved truncation errors as low as 10$^{-7}$ or better. 

For molecules in the test dataset, the active spaces were reconstructed by selecting $p_z$ orbitals of carbon and nitrogen atoms following the split-localization procedure. This process yielded the following CAS configurations: I -- (22e, 22o), II -- (32e, 32o), III -- (40e, 40o), IV -- (44e, 44o), and V -- (36e, 33o).

\begin{figure*}[!ht]
  \subfloat[\label{fig_I}]{%
    \includegraphics[width=6cm]{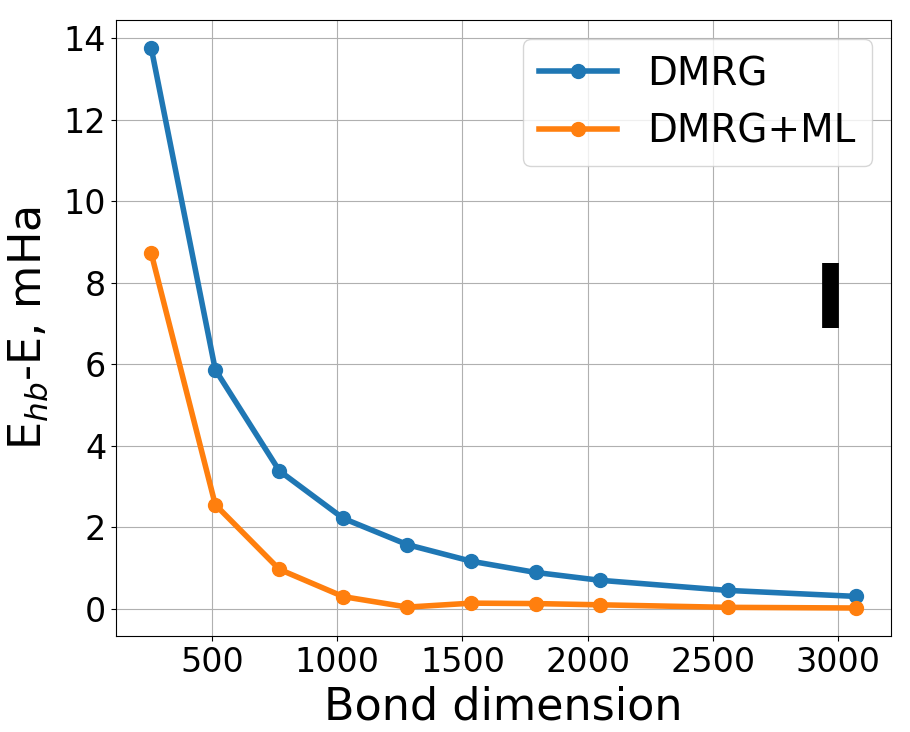}
  }
  \subfloat[\label{fig_II}]{%
    \includegraphics[width=6cm]{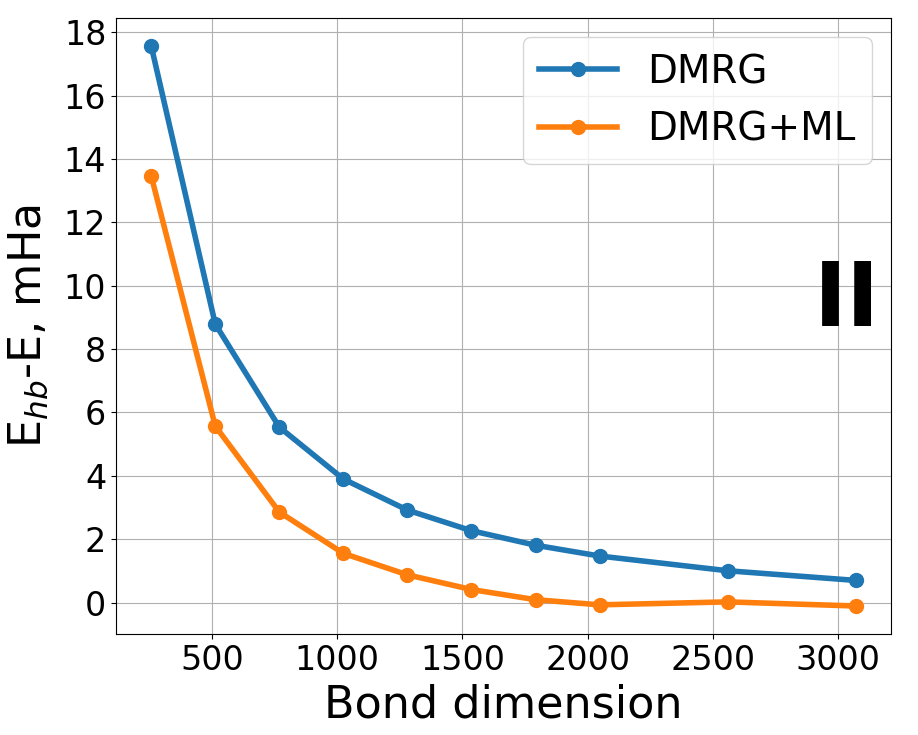}
  }
  \subfloat[\label{fig_III}]{%
    \includegraphics[width=6cm]{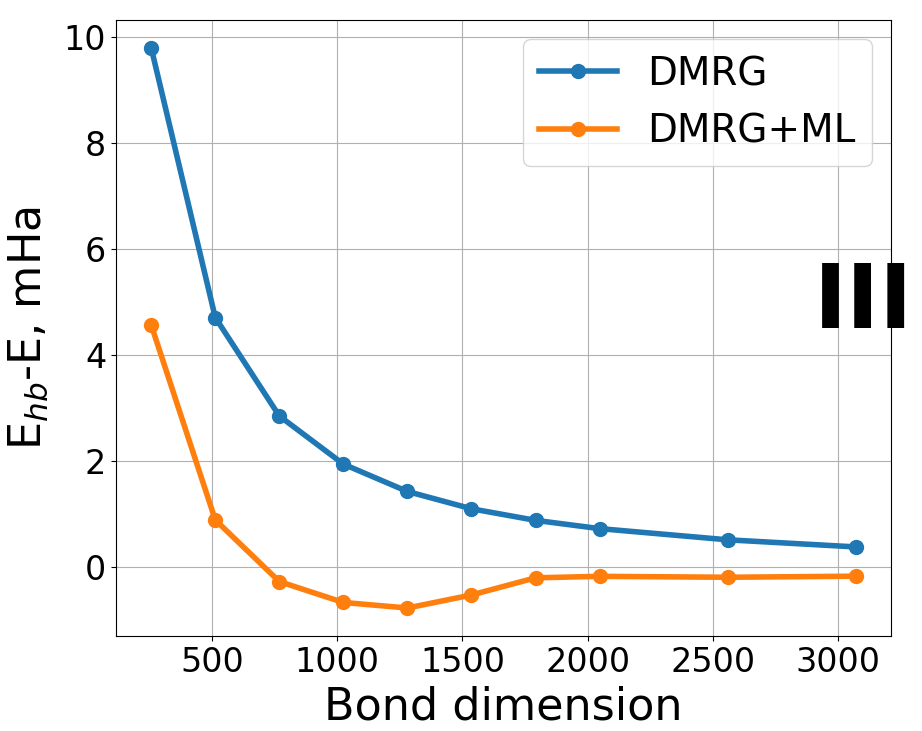}
  } \\
  \subfloat[\label{fig_V}]{%
    \includegraphics[width=6cm]{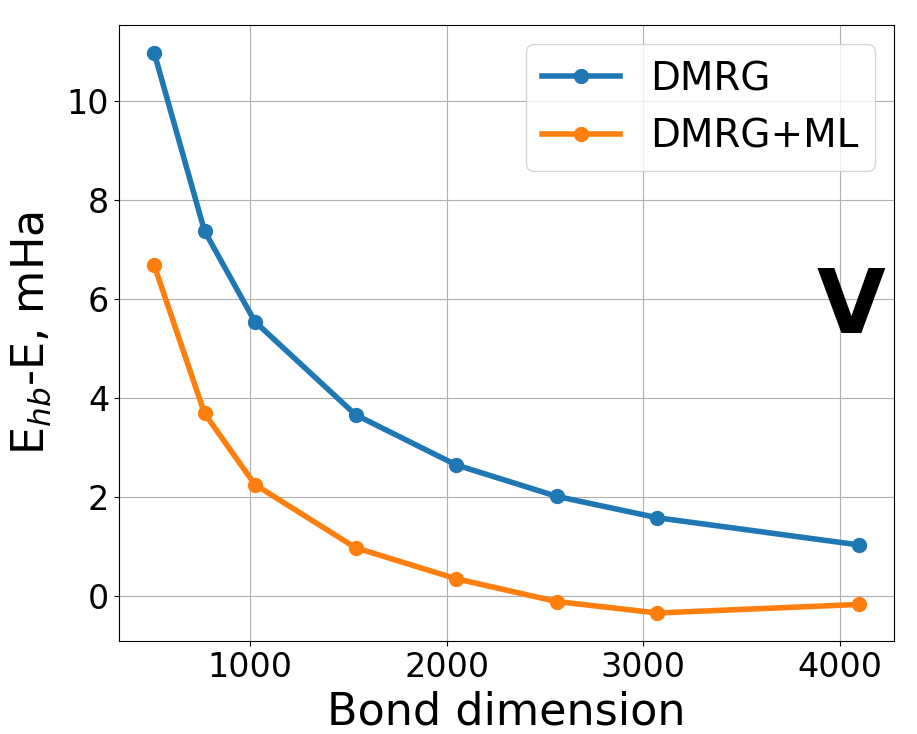}
  }
  \subfloat[\label{fig_TRE}]{%
    \includegraphics[width=6cm]{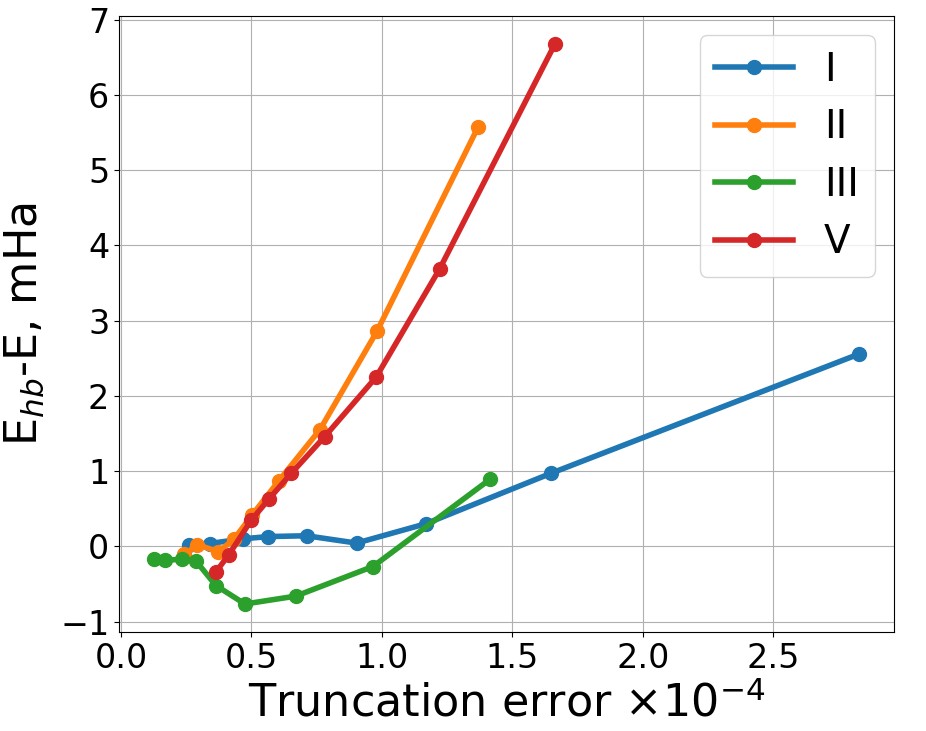}
  }
  \caption{(a–d) Differences between reference (high-bond-dimension, $E_{hb}$) DMRG energies and standard DMRG energies, as well as ML-corrected DMRG energies, calculated across different bond dimensions for the singlet states of molecules shown in Figure \ref{fig:test_mols}. (e) Dependence of DMRG+ML energies on the truncation error of the parent low-bond-dimension DMRG calculations for all four molecular examples.}
  \label{fig:test_mols_en_diff}
\end{figure*}

The results of our $\Delta$-ML DMRG approach, compared to standard DMRG, for the lowest singlet state energies of PAHs \textbf{I}, \textbf{II}, \textbf{III}, and \textbf{V} from the test set are shown in Figure \ref{fig:test_mols_en_diff}. As illustrated, the ML-corrected DMRG consistently provides lower (and occasionally very slightly overshot) energies that are more accurate than the standard DMRG.

At very low bond dimensions, the ML corrections do not yet achieve chemical accuracy (1 kcal/mol). At these bond dimensions, the results are insufficiently accurate to serve as reliable input for the model. However, as the bond dimension increases, the quality of the ML corrections improves. For bond dimensions between 750 and 1250, the difference between the ML-corrected energies and $E_{hb}$ energies consistently falls below 1 mHa, as seen for all test molecules. The performance of the ML model is further validated as the bond dimension approaches the high-quality limit, where the ML corrections maintain accuracy without overestimating the energies.

Figure \ref{fig_TRE} shows that, at TRE of $5 \times 10^{-5}$ (system \textbf{II}), the accuracy of the ML corrections reaches 1 mHa in the worst case. For the other examples, this level of accuracy is achieved at higher truncation errors.

A particularly challenging test for the ML model is to correctly predict energy gaps between close-lying electronic states. Depending on their size, geometry, and edge structure, peri-fused PAHs can adopt either 
closed-shell singlet, open-shell singlet, or higher-spin ground states, with the energy difference between them often being very small.
Diradical neutral [3]trianguelene (\textbf{I}) is known to be the smallest PAH with triplet ground state. Owing to the maximum overlap of the $\pi$-radical wave functions it is considered to have the largest singlet-triplet (S-T) gap among all PAH diradicals.\cite{Ortiz-Boto-2019} In contrast, peripentacene (\textbf{IV}) exhibits a singlet open-shell ground state. Its S-T gap has been estimated to be around 130 meV from all-$\pi$ DMRG calculations, and approximately 49 meV after accounting for the substrate effects and dynamical correlation corrections via multireference-coupled cluster methods.\cite{Grande-Urgel-2020}

\begin{figure}[!ht]
\includegraphics[width=8cm]{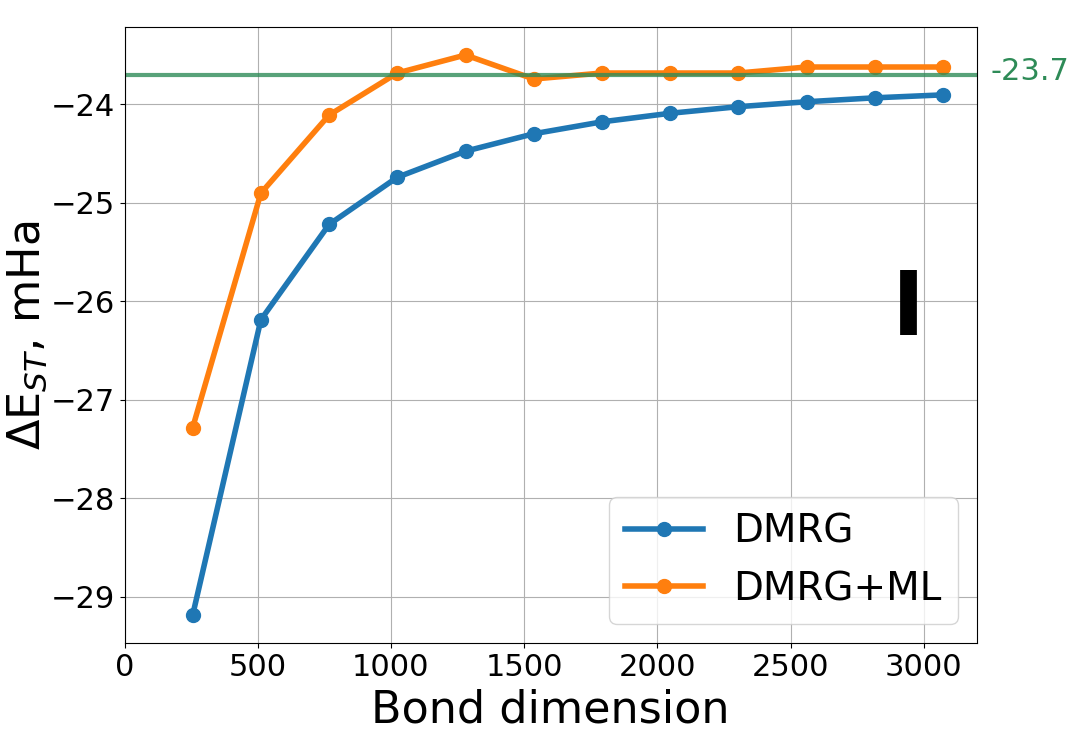}
\caption{Singlet-triplet (S-T) gap for [3]triangulene (\textbf{I}) computed at various bond dimensions. The reference S-T gap value of -2.375$\times$10$^{-2}$ Ha is obtained from a DMRG calculation with a bond dimension of 5000.}
\label{fig:triangulene_st_gap}
\centering
\end{figure}

\begin{figure}[!ht]
\includegraphics[width=8cm]{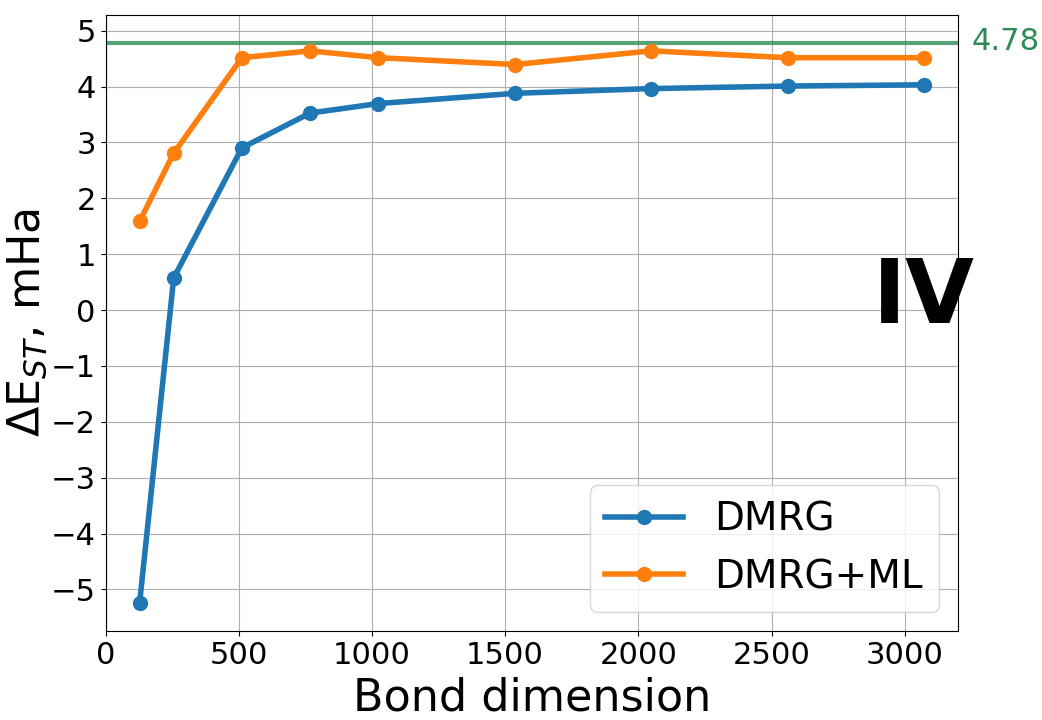}
\caption{Singlet-triplet (S-T) gap for peripentacene (\textbf{IV}) computed at various bond dimensions. The reference S-T gap value of 4.78$\times$10$^{-3}$ Ha is taken from S\'{a}nchez-Grande et al\cite{Grande-Urgel-2020}.}
\label{fig:perip_st_gap}
\centering
\end{figure}

In Figure \ref{fig:triangulene_st_gap}, we present the performance of the ML corrected DMRG on the S-T gap of [3]triangulene. It is evident that our $\Delta$-ML DMRG approach correctly predicts the S-T gap value already at $M=750$, which corresponds to TRE in the range of 1--3$\times$10$^{-4}$. A slight deviation is observed with increase of the bond dimension, however, this deviation never exceeds 1 mHa in absolute value. 

Figure \ref{fig:perip_st_gap} illustrates the S-T gap results for peripentacene. In this case, standard DMRG even fails to predict the correct ground state at low bond dimensions. In contrast, the $\Delta$-ML DMRG method predicts the correct order of both electronic states much earlier, already at $M=500$ and TRE of 1$\times$10$^{-4}$. For bond dimensions greater than 500, $\Delta$-ML DMRG shows stable predictions within the range 4.5--4.7$\times$10$^{-3}$ Ha, closely aligning with the reference value of 4.78$\times$10$^{-3}$ Ha, obtained by S\'{a}nchez-Grande et al.\cite{Grande-Urgel-2020} after extrapolation of DMRG energies at high bond dimensions with respect to TRE. 
This remarkable agreement is particularly promising, especially considering that, despite selecting PAHs from the COMPAS-1D database with the smallest HOMO-LUMO gaps, none of the systems in the training set exhibited an open-shell singlet ground state like that of peripentacene.

In summary, we demonstrated the potential of a simple ML model to significantly enhance the performance of the quantum chemical DMRG method. The model leverages minimal input from low-cost DMRG calculations -- such as single- and two-site entropies, truncation error, and orbital occupancies -- without requiring prior knowledge of molecular geometry or other properties. This approach substantially improves the accuracy of low-cost DMRG energies and energy gaps. We applied the model to the electronic structure of polycyclic aromatic hydrocarbons, making the trained model, available online \cite{github}, directly applicable to general organic aromatic $\pi$-extended systems, including aromatic heterocycles.

Furthermore, we believe similar models can be developed for other classes of strongly correlated molecules, such as transition metal complexes \cite{Roemelt2019}. To support this, we advocate for the systematic tabulation and sharing of DMRG calculation results.

\begin{acknowledgement}
This material is based upon work supported by the U.S. Department of Energy, Office of Science, Office of Advanced Scientific Computing Research and Office of Basic Energy Sciences, Scientific Discovery through Advanced Computing (SciDAC) program under Award Number DE-SC0022198.  This research used resources of the National Energy Research Scientific Computing Center, a DOE Office of Science User Facility supported by the Office of Science of the U.S. Department of Energy under Contract No. DE-AC02-05CH11231 using NERSC award BES-ERCAP0029462.

L.V. further acknowledges support from the Czech Science Foundation (grant no. 23-05486S), the Ministry of
Education, Youth and Sports of the Czech Republic through the e-INFRA CZ (ID:90254), and the Advanced Multiscale Materials for Key Enabling Technologies project, supported by the Ministry of Education, Youth, and Sports of the Czech Republic. Project No. CZ.02.01.01/00/22\_008/0004558, Co-funded by the European Union.

\end{acknowledgement}

%\begin{suppinfo}

%Sup

%\end{suppinfo}

\bibliography{references}

\end{document}

\section{Computational and ML model details}

Input orbitals for DMRG calculations were computed at the DFT level using the B3LYP exchange-correlation functional \cite{Parr88_785,Becke88_3098} and the cc-PVDZ basis set \cite{Dunning1989}. To improve the performance of DMRG and facilitate active space selection the split-localization procedure (separate localization
of occupied and virtual orbitals) by means of the   Pipek-Mezey method \cite{Pipek1989} was used. Complete $\pi$-active spaces were selected afterward, leading to the following active space sizes -- (22e, 22o), (26e, 26o), (30e, 30o) for molecules with 5, 6, and 7 benzene rings (5C, 6C, and 7C) respectively. The ordering of the active orbitals mapped onto the 1D lattice of DMRG sites was obtained with the Fiedler method \cite{Barcza2011} applied on the matrix of exchange integrals \cite{Olivares-Amaya-Hu-2015}. All DMRG calculations were initialized with the CI-DEAS procedure \cite{Legeza-Solyom-2003}.

DMRG calculations with bond dimensions 64, 128, and 256 for 5C and 6C molecules, and with bond dimensions 128, 256, and 512 for 7C molecules were used as training points. This ensured the input data with the truncation error in the 1$\times$10$^{-4}$ -- 5$\times$10$^{-6}$ range. The reference targets were obtained with bond dimension 3000 for 5C and 6C molecules, and with bond dimension 4000 for 7C molecules. The reference accurate calculations ensured the truncation error ranged from 10$^{-7}$ to lower orders.

Test molecules were subject to the same active space selection procedure, that resulted in the following active space configurations: \textbf{I} -- (22e, 22o), \textbf{II} -- (32e, 32o), \textbf{III} -- (40e, 40o), \textbf{IV} -- (44e, 44o), \textbf{V} -- (36e, 33o).

Message passing graph neural network (MPGNN) was chosen as a machine learning model. 
%In case of MPGNN, the node representation in each layer $k$ is typically expressed as 
%\begin{equation}
% N^{k}_{i} = g \left( N^{k-1}_{i}, \underset{j \in \text{conn}} {\bigoplus} f \left( N^{k-1}_{i}, N^{k-1}_{j}, E_{ij} \right) \right),
%\end{equation}
%\noindent
%where $N^{k}_{i}$ denote $i$th node features in $k$th layer, $E_{ij}$ represent edge features and $f$ and $g$ are general differentiable functions (e.g. neural networks).
Within the MPGNN framework for each node an embedding is built based on its connections. To briefly introduce the architecture of the MPGNN employed in this work, consider node 1 with node features ${N^{k-1}_1}$ from ($k-1$)th massage passing layer, and connected to nodes 2 and 3 with node features ${N^{k-1}_2}$, ${N^{k-1}_3}$ respectively, and edge features ${E_{12}}$, ${E_{13}}$. Initially, messages from the connected neighbors are generated by concatenating their node and edge features and processing them using a differentiable function $f$ (e.g. fully connected neural network)

\begin{eqnarray}
  M^k_{12} & = & f(N_1^{k-1}, N_2^{k-1}, E_{12}), \\
  M^k_{13} & = & f(N_1^{k-1}, N_3^{k-1}, E_{13}).
\end{eqnarray}

\noindent
Next, the messages from the connected neighbors are aggregated using an operation $\bigoplus$ (e.g., summation or mean) as $\bigoplus \left(M^k_{12}, M^k_{13}\right)$ and concatenated with the node features from the ($k-1$)th massage passing layer, ${N^{k-1}_1}$.
This combined vector is then passed through another set of fully connected neural network layers (referred to as the message update), function $g$, resulting in updated node features ${N^{k}_1}$ for the $k$th massage passing layer. 
By applying multiple message passing layers, feature representations for all nodes are obtained, capturing their specific interconnections. Finally, a feature representation of the entire graph can be constructed by applying an aggregation function to the final feature vectors of all graph nodes.

The simplest model (\textbf{model1}), whose results are presented in the main text, assumed one message passing layer with $f$ and $g$ being identity functions. The second model (\textbf{model2}), whose results are presented below, featured three message passing layers.
In this model, the function $f$ and $g$ were single-layer fully connected neural networks with 8 neurons each, using ReLU activation ($f(z) = max \{ 0,z \}$) for each neuron.
Both models featured mean aggregation function. Resulted graph representation was processed through tree-layer fully-connected neural network, with 20 neurons in each layer and leaky ReLU ($f(z) = max \{ 0,z \} + 0.2 \cdot min \{ 0,z \}$) activation function on each neuron. To avoid overfitting dropout regularization\cite{Hinton-Srivastava-at-al-2012, Srivastava-Hinton-at-al-2014} procedure was used with dropout rate 0.1. Each train procedure run over 1500 epochs.

One-site entropy was used as a node feature, while mutual information served as an edge feature. Additionally, orbital occupation information, encoded as a three-element one-hot vector (\textit{occup}), was incorporated into the edge features. If both nodes were associated with occupied orbitals, then $occup[0]=1$. If both were associated with virtual orbitals, $occup[2]=1$. Otherwise, $occup[1]=1$.

TensorFlow machine learning platform\cite{tensorflow} and PyTorch Geometric library\cite{pytorch_geom} were used to built the models.

\section{Geometries}

Geometries of the test set molecules are provided in Tables \ref{tab:s_triangulene_xyz} - \ref{tab:s_V_xyz}. 

\begin{table}[!ht]
\begin{scriptsize}
\begin{center}
\begin{tabular}{c c c c}
C  & -4.4227  & -0.7180  & -0.0001 \\
H  & -5.4888  & -0.8911  &  0.0008 \\
C  & -3.9178  &  0.6309  & -0.0006 \\
H  & -4.6136  &  1.4569  & -0.0003 \\
C  & -2.4184  &  0.9156  & -0.0011 \\
C  & -1.9322  &  2.3672  & -0.0018 \\
H  & -2.6151  &  3.2039  & -0.0025 \\
C  & -0.4126  &  2.5527  & -0.0009 \\
C  &  0.1668  &  3.9646  & -0.0005 \\
H  & -0.5030  &  4.8118  & -0.0015 \\
C  &  1.5895  &  4.1891  &  0.0018 \\
H  &  1.9727  &  5.1989  &  0.0034 \\
C  &  2.5052  &  3.0773  &  0.0025 \\
H  &  3.5685  &  3.2669  &  0.0049 \\
C  &  2.0020  &  1.6365  &  0.0005 \\
C  &  0.5004  &  1.3190  &  0.0000 \\
C  & -0.0001  & -0.0001  & -0.0004 \\
C  &  0.8921  & -1.0930  & -0.0006 \\
C  &  2.4170  & -0.9189  & -0.0010 \\
C  &  3.3503  & -2.1265  & -0.0007 \\
H  &  4.4189  & -1.9699  & -0.0008 \\
C  &  2.8334  & -3.4709  &  0.0003 \\
H  &  3.5164  & -4.3076  &  0.0010 \\
C  &  1.4127  & -3.7082  &  0.0010 \\
H  &  1.0454  & -4.7238  &  0.0026 \\
C  &  0.4162  & -2.5522  & -0.0007 \\
C  & -1.0840  & -2.8570  & -0.0012 \\
H  & -1.4671  & -3.8668  & -0.0013 \\
C  & -2.0045  & -1.6338  & -0.0009 \\
C  & -1.3926  & -0.2262  & -0.0014 \\
C  & -3.5170  & -1.8380  &  0.0052 \\
H  & -3.9158  & -2.8416  &  0.0097 \\
C  &  3.0162  &  0.4898  & -0.0006 \\
H  &  4.0822  &  0.6629  & -0.0028

\end{tabular}
\caption{Geometry of the system \textbf{I} ([3]triangulene]) in XYZ format (\AA).}
\label{tab:s_triangulene_xyz}
\end{center}
\end{scriptsize}
\end{table}

\begin{table}[!ht]
\begin{scriptsize}
\begin{center}
\begin{tabular}{c c c c}
C  &  0.0000  & -0.7061  & -0.0001 \\
C  &  0.0000  &  0.7061  & -0.0001 \\
C  &  1.2234  & -1.4132  & -0.0001 \\
C  &  1.2233  &  1.4131  &  0.0000 \\
C  & -1.2234  & -1.4132  &  0.0000 \\
C  & -1.2233  &  1.4132  &  0.0000 \\
C  &  2.4487  & -0.7068  & -0.0002 \\
C  &  2.4486  &  0.7068  & -0.0001 \\
C  & -2.4487  & -0.7068  &  0.0000 \\
C  & -2.4486  &  0.7069  & -0.0001 \\
C  &  1.2227  & -2.8249  &  0.0001 \\
C  &  1.2228  &  2.8249  &  0.0002 \\
C  & -1.2228  & -2.8249  &  0.0001 \\
C  & -1.2227  &  2.8249  &  0.0000 \\
C  &  3.6724  & -1.4131  & -0.0002 \\
C  &  3.6723  &  1.4130  & -0.0001 \\
C  & -3.6724  & -1.4131  &  0.0000 \\
C  & -3.6723  &  1.4131  & -0.0001 \\
C  &  0.0000  & -3.5128  &  0.0002 \\
C  &  0.0001  &  3.5128  &  0.0002 \\
C  &  2.4451  & -3.5102  &  0.0001 \\
C  &  2.4451  &  3.5102  &  0.0003 \\ 
C  & -2.4451  & -3.5102  &  0.0003 \\
C  & -2.4450  &  3.5102  &  0.0000 \\
C  &  3.6544  & -2.8137  & -0.0001 \\
C  &  3.6544  &  2.8136  &  0.0002 \\
C  & -3.6545  & -2.8136  &  0.0002 \\
C  & -3.6544  &  2.8137  & -0.0001 \\
C  &  4.8770  & -0.6978  & -0.0003 \\
C  &  4.8771  &  0.6978  &  0.0000 \\
C  & -4.8771  & -0.6977  & -0.0001 \\
C  & -4.8771  &  0.6978  & -0.0002 \\
H  &  0.0000  & -4.6028  &  0.0002 \\
H  &  0.0001  &  4.6029  &  0.0004 \\
H  &  2.4693  & -4.5981  &  0.0002 \\
H  &  2.4694  &  4.5981  &  0.0005 \\
H  & -2.4694  & -4.5981  &  0.0004 \\
H  & -2.4693  &  4.5981  &  0.0000 \\
H  &  4.5838  & -3.3794  &  0.0000 \\
H  &  4.5839  &  3.3793  &  0.0004 \\
H  & -4.5839  & -3.3794  &  0.0003 \\
H  & -4.5838  &  3.3794  & -0.0002 \\
H  &  5.8313  & -1.2208  & -0.0003 \\
H  &  5.8313  &  1.2207  &  0.0000 \\
H  & -5.8313  & -1.2206  & -0.0001 \\
H  & -5.8313  &  1.2208  & -0.0002
\end{tabular}
\caption{Geometry of the system \textbf{II} (ovalene) in XYZ format (\AA).}
\label{tab:s_ovalene_xyz}
\end{center}
\end{scriptsize}
\end{table}

\begin{table}[!ht]
\begin{scriptsize}
\begin{center}
\begin{tabular}{c c c c}
    C  &  1.2470  &  0.7040  & -0.0009 \\
    C  &  1.2470  & -0.7041  & -0.0007 \\
    C  & -1.2470  &  0.7040  & -0.0010 \\
    C  & -1.2471  & -0.7042  & -0.0012 \\
    C  &  0.0000  &  1.3979  & -0.0010 \\
    C  &  0.0000  & -1.3980  & -0.0009 \\
    C  &  2.4797  &  1.4156  & -0.0008 \\
    C  &  2.4798  & -1.4156  &  0.0000 \\
    C  & -2.4797  &  1.4155  & -0.0004 \\
    C  & -2.4797  & -1.4157  & -0.0015 \\
    C  &  0.0001  &  2.8084  & -0.0013 \\
    C  &  0.0000  & -2.8085  & -0.0007 \\ 
    C  &  3.7271  &  0.7078  & -0.0003 \\
    C  &  3.7271  & -0.7078  &  0.0006 \\
    C  & -3.7270  &  0.7077  & -0.0001 \\
    C  & -3.7270  & -0.7079  & -0.0015 \\
    C  &  2.4065  &  2.8256  & -0.0010 \\
    C  &  2.4066  & -2.8257  &  0.0001 \\
    C  & -2.4064  &  2.8256  & -0.0007 \\
    C  & -2.4066  & -2.8257  & -0.0010 \\
    C  &  1.1976  &  3.5096  & -0.0014 \\
    C  & -1.1975  &  3.5095  & -0.0011 \\
    C  &  1.1976  & -3.5097  & -0.0002 \\
    C  & -1.1975  & -3.5097  & -0.0005 \\
    C  &  4.9896  &  1.3690  & -0.0007 \\
    C  &  4.9898  & -1.3690  &  0.0018 \\
    C  & -4.9896  &  1.3690  &  0.0018 \\
    C  & -4.9897  & -1.3690  & -0.0028 \\
    C  &  6.2236  &  0.6985  &  0.0005 \\
    C  &  6.2238  & -0.6984  &  0.0019 \\
    C  & -6.2237  &  0.6985  &  0.0019 \\
    C  & -6.2238  & -0.6984  & -0.0008 \\
    C  &  7.4495  &  1.3891  &  0.0000 \\
    C  &  7.4498  & -1.3889  &  0.0029 \\
    C  & -7.4496  &  1.3892  &  0.0047 \\
    C  & -7.4498  & -1.3888  & -0.0014 \\
    C  &  8.6595  &  0.6954  &  0.0004 \\
    C  &  8.6595  & -0.6950  &  0.0019 \\
    C  & -8.6595  &  0.6954  &  0.0043 \\
    C  & -8.6596  & -0.6949  &  0.0012 \\
    H  &  3.2969  &  3.4475  & -0.0009 \\
    H  &  3.2969  & -3.4475  &  0.0004 \\
    H  & -3.2968  &  3.4475  & -0.0007 \\
    H  & -3.2969  & -3.4477  & -0.0008 \\
    H  &  1.2130  &  4.5975  & -0.0015 \\
    H  & -1.2129  &  4.5975  & -0.0012 \\
    H  &  1.2131  & -4.5976  &  0.0000 \\
    H  & -1.2129  & -4.5976  & -0.0001 \\
    H  &  5.0376  &  2.4561  & -0.0018 \\
    H  &  5.0378  & -2.4562  &  0.0027 \\
    H  & -5.0375  &  2.4562  &  0.0037 \\
    H  & -5.0377  & -2.4561  & -0.0048 \\
    H  &  7.4699  &  2.4770  & -0.0012 \\
    H  &  7.4703  & -2.4768  &  0.0040 \\
    H  & -7.4700  &  2.4771  &  0.0071 \\
    H  & -7.4704  & -2.4767  & -0.0035 \\
    H  &  9.5987  &  1.2405  & -0.0002 \\ 
    H  &  9.5988  & -1.2399  &  0.0024 \\
    H  & -9.5987  &  1.2405  &  0.0063 \\
    H  & -9.5989  & -1.2398  &  0.0009
\end{tabular}
\caption{Geometry of the system \textbf{III} in XYZ format (\AA).}
\label{tab:s_III_xyz}
\end{center}
\end{scriptsize}
\end{table}

\begin{table}[!ht]
\label{tab:s_peripentacene_xyz}
\begin{scriptsize}
\begin{center}
\begin{tabular}{c c c c}
C & 14.13369621 & 24.09738288 & 7.41419320 \\
C & 15.52860709 & 24.11671801 & 7.41418866 \\
C & 16.27758770 & 22.93689028 & 7.41418564 \\
C & 17.74444497 & 22.93682197 & 7.41418441 \\
C & 18.49325442 & 24.11677618 & 7.41418346 \\
C & 19.88806712 & 24.09763674 & 7.41418100 \\
C & 20.57223729 & 22.90243977 & 7.41418259 \\
C & 19.86834535 & 21.67999544 & 7.41418577 \\
C & 20.54741789 & 20.44587349 & 7.41419006 \\
C & 19.86832179 & 19.22959736 & 7.41419500 \\
C & 20.54839156 & 17.99915476 & 7.41420924 \\
C & 19.86812585 & 16.76892065 & 7.41421238 \\
C & 20.54707494 & 15.55233190 & 7.41419817 \\
C & 19.86777884 & 14.31850274 & 7.41419387 \\
C & 18.43606199 & 14.30213245 & 7.41419137 \\
C & 20.57149434 & 13.09598297 & 7.41419271 \\
C & 19.88719903 & 11.90094885 & 7.41418828 \\
C & 18.49228589 & 11.88199236 & 7.41418659 \\
C & 17.74374134 & 13.06205824 & 7.41419023 \\
C & 16.27683440 & 13.06230394 & 7.41418831 \\
C & 15.52769994 & 11.88255792 & 7.41418569 \\
C & 14.13277939 & 11.90212127 & 7.41418335 \\
C & 13.44907613 & 13.09744558 & 7.41418412 \\
C & 14.15343034 & 14.31959321 & 7.41418731 \\
C & 15.58516328 & 14.30264567 & 7.41418791 \\
C & 16.29582174 & 15.54366728 & 7.41419103 \\
C & 17.72594899 & 15.54346419 & 7.41419284 \\
C & 18.42404082 & 16.76403810 & 7.41419318 \\
C & 17.72016424 & 17.99945327 & 7.41419368 \\
C & 18.42423720 & 19.23472914 & 7.41419295 \\
C & 17.72628478 & 20.45532510 & 7.41418692 \\
C & 16.29617552 & 20.45553751 & 7.41418803 \\
C & 18.43657439 & 21.69663318 & 7.41418555 \\
C & 15.58570198 & 21.69660930 & 7.41418551 \\
C & 14.15398921 & 21.67992894 & 7.41418556 \\
C & 13.44979816 & 22.90219735 & 7.41419266 \\
C & 13.47500551 & 20.44609659 & 7.41418861 \\
C & 14.15433553 & 19.22994299 & 7.41419101 \\
C & 15.59828148 & 19.23486604 & 7.41419063 \\
C & 15.59809424 & 16.76441229 & 7.41419457 \\
C & 16.30229099 & 17.99958278 & 7.41419044 \\
C & 14.15411872 & 16.76953807 & 7.41418855 \\
C & 13.47400409 & 17.99979206 & 7.41417884 \\
C & 13.47461040 & 15.55350292 & 7.41418763 \\
H & 13.59028745 & 25.03198133 & 7.41419659 \\
H & 16.02125699 & 25.07598052 & 7.41418836 \\
H & 18.00014455 & 25.07580095 & 7.41418311 \\
H & 20.43145300 & 25.03222835 & 7.41417890 \\
H & 21.65369521 & 22.88888590 & 7.41418195 \\
H & 21.62972366 & 20.44627211 & 7.41419173 \\
H & 21.62938053 & 15.55201132 & 7.41420002 \\
H & 21.65294363 & 13.10947409 & 7.41419432 \\
H & 20.43032253 & 10.96620056 & 7.41418719 \\
H & 17.99917719 & 10.92296453 & 7.41418390 \\
H & 16.02022255 & 10.92323413 & 7.41418572 \\
H & 13.58913150 & 10.96767614 & 7.41418106 \\
H & 12.36763995 & 13.11178004 & 7.41418273 \\
H & 12.36836644 & 22.88799347 & 7.41419633 \\
H & 12.39272858 & 20.44587882 & 7.41418770 \\
H & 12.39183423 & 17.99987347 & 7.41417405 \\
H & 12.39233706 & 15.55384728 & 7.41418436 \\
H & 21.63056990 & 17.99921526 & 7.41421445
\end{tabular}
\caption{Geometry of the system \textbf{IV} (peripentacene) in XYZ format (\AA).}
\end{center}
\end{scriptsize}
\end{table}

\begin{table}[!ht]
\begin{scriptsize}
\begin{center}
\begin{tabular}{c c c c}
    C  & -5.6025  & -0.0066  &  0.0009 \\
    H  & -6.6825  & -0.0079  &  0.0020 \\
    C  & -4.9125  &  1.1916  & -0.0006 \\
    H  & -5.4544  &  2.1258  & -0.0010 \\
    C  & -3.5092  &  1.1956  & -0.0009 \\
    C  & -2.8518  &  2.4343  & -0.0019 \\
    H  & -3.4258  &  3.3491  & -0.0024 \\
    C  & -1.4604  &  2.4870  & -0.0016 \\
    C  & -0.6970  &  3.6621  & -0.0015 \\
    H  & -1.1886  &  4.6237  & -0.0020 \\
    C  &  0.6929  &  3.5934  & -0.0001 \\
    C  &  1.4606  &  4.7782  &  0.0008 \\
    H  &  0.9820  &  5.7464  &  0.0011 \\
    C  &  2.8214  &  4.6725  &  0.0017 \\
    H  &  3.4204  &  5.5712  &  0.0028 \\
    N  &  3.4255  &  3.4908  &  0.0016 \\
    C  &  2.7423  &  2.3440  &  0.0008 \\
    N  &  1.3691  &  2.3902  & -0.0001 \\
    C  &  0.6577  &  1.2137  & -0.0008 \\
    C  &  1.3508  &  0.0016  & -0.0007 \\
    C  &  2.8214  &  0.0034  &  0.0001 \\
    N  &  3.4286  & -1.1818  &  0.0002 \\
    C  &  2.7481  & -2.3373  & -0.0002 \\
    N  &  1.3750  & -2.3870  & -0.0006 \\
    C  &  0.6607  & -1.2123  & -0.0009 \\
    C  & -0.7338  & -1.2048  & -0.0010 \\
    C  & -1.4500  & -0.0018  & -0.0007 \\
    N  & -2.8257  & -0.0034  &  0.0000 \\
    C  & -3.5064  & -1.2041  & -0.0001 \\
    C  & -2.8459  & -2.4412  & -0.0010 \\
    H  & -3.4178  & -3.3574  & -0.0012 \\
    C  & -1.4544  & -2.4908  & -0.0010 \\
    C  & -0.6879  & -3.6638  & -0.0006 \\
    H  & -1.1772  & -4.6267  & -0.0012 \\
    C  &  0.7018  & -3.5919  & -0.0002 \\
    C  &  1.4726  & -4.7746  &  0.0003 \\
    H  &  0.9964  & -5.7440  &  0.0001 \\
    C  &  2.8331  & -4.6655  &  0.0007 \\
    H  &  3.4343  & -5.5627  &  0.0008 \\
    N  &  3.4343  & -3.4823  &  0.0001 \\
    C  & -4.9097  & -1.2032  &  0.0068 \\
    H  & -5.4494  & -2.1387  &  0.0071 \\
    C  & -0.7368  &  1.2028  & -0.0011 \\
    N  &  3.4257  &  1.1902  &  0.0008
\end{tabular}
\caption{Geometry of the system \textbf{V} in XYZ format (\AA).}
\label{tab:s_V_xyz}
\end{center}
\end{scriptsize}
\end{table}

\section{Additional results}

\begin{figure*}[!ht]
  \subfloat[\label{fig_I}]{%
    \includegraphics[width=6cm]{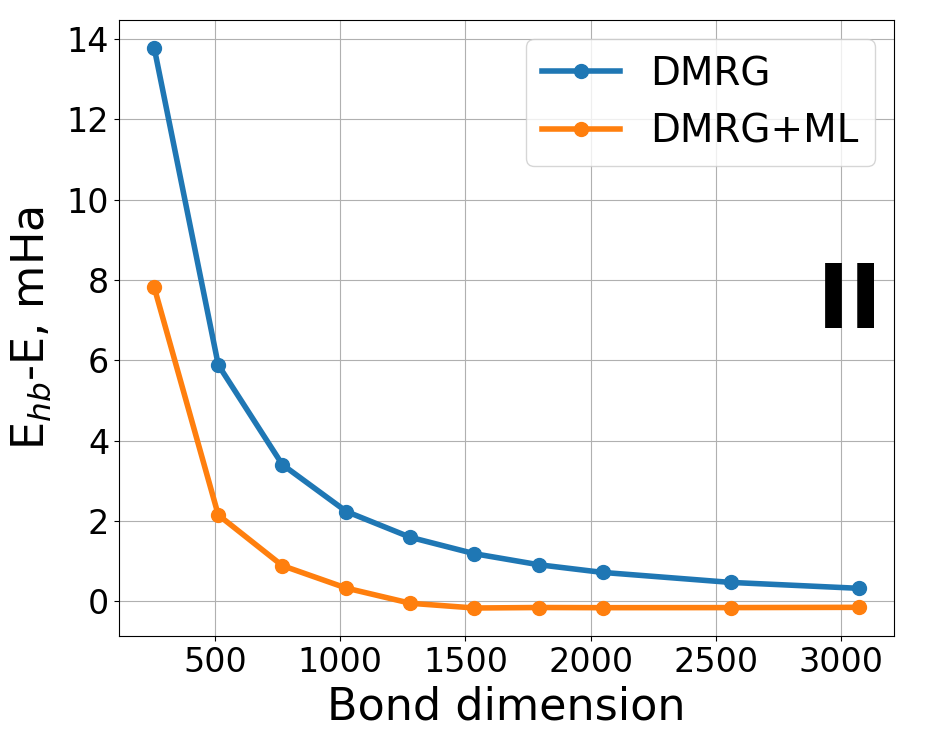}
  }
  \subfloat[\label{fig_II}]{%
    \includegraphics[width=6cm]{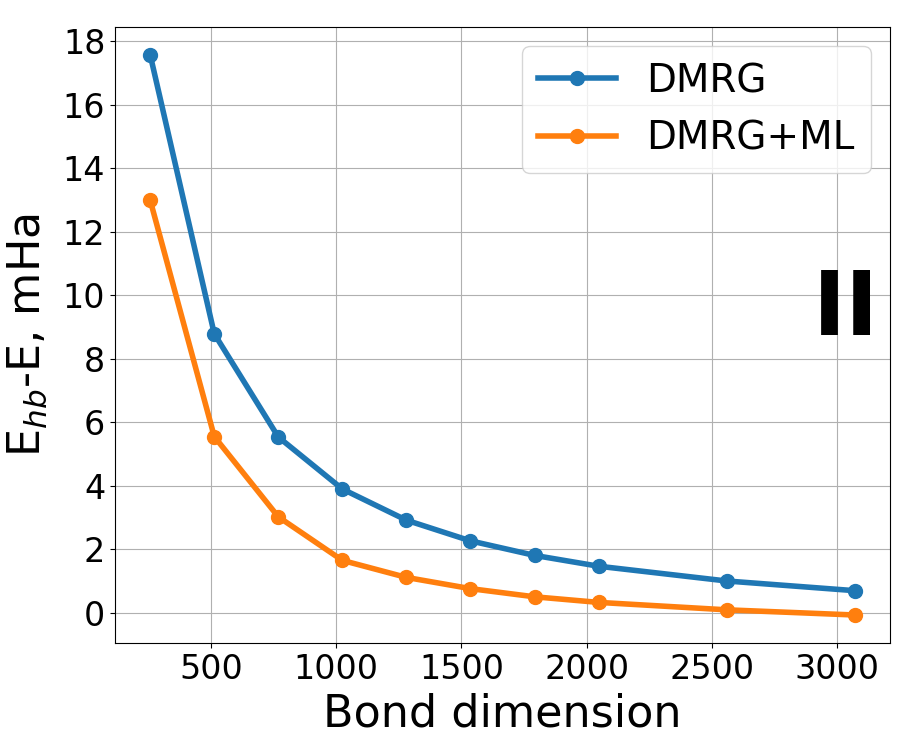}
  }
  \subfloat[\label{fig_III}]{%
    \includegraphics[width=6cm]{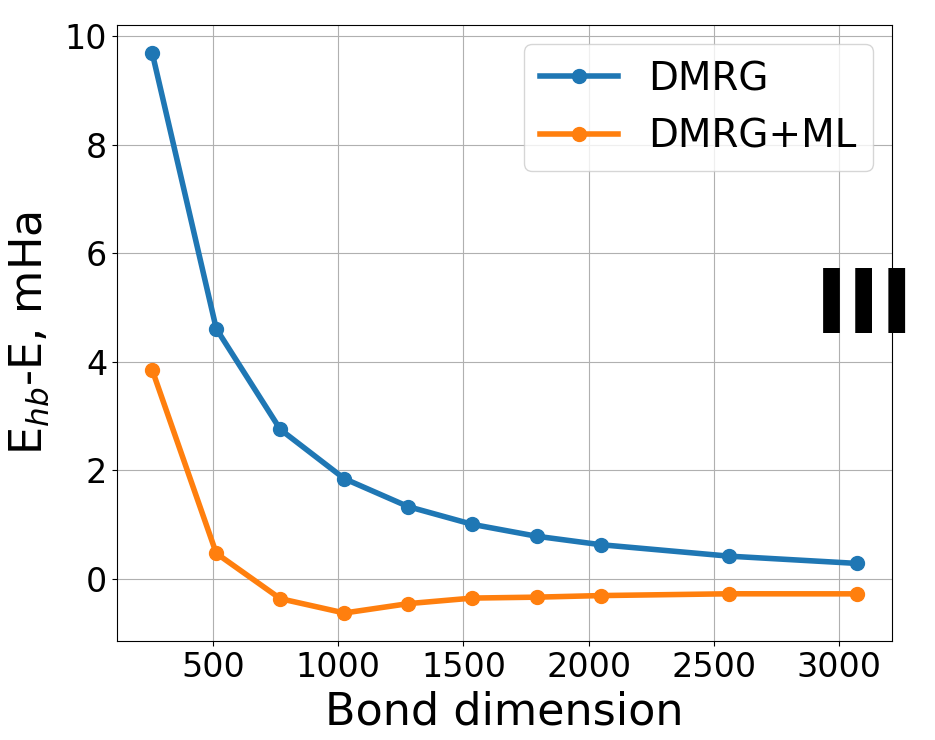}
  } \\
  \subfloat[\label{fig_V}]{%
    \includegraphics[width=6cm]{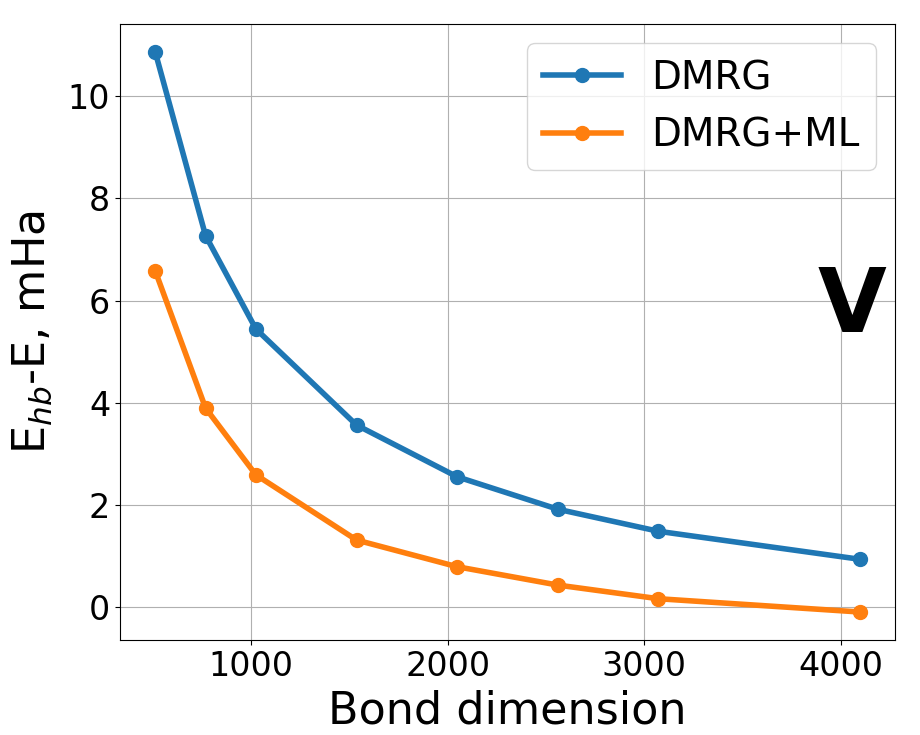}
  }
  \subfloat[\label{fig_TRE}]{%
    \includegraphics[width=6cm]{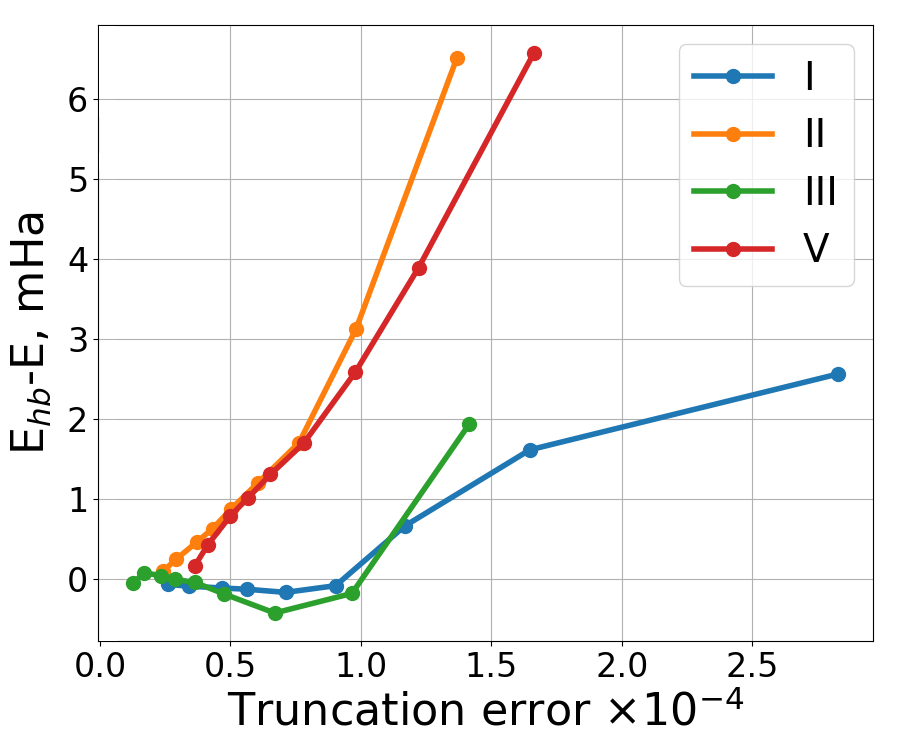}
  }
  \caption{(a–d) Differences between reference (high-bond-dimension, $E_{hb}$) DMRG energies and standard DMRG energies, as well as ML-corrected DMRG energies, calculated across different bond dimensions for the singlet states of molecules shown in Figure 2 of the main text. (e) Dependence of DMRG+ML energies on the truncation error of the parent low-bond-dimension DMRG calculations for all four molecular examples. \textbf{model2} predictions.}
  \label{fig:test_mols_en_diff}
\end{figure*}

\begin{figure}[h]
\includegraphics[width=12cm]{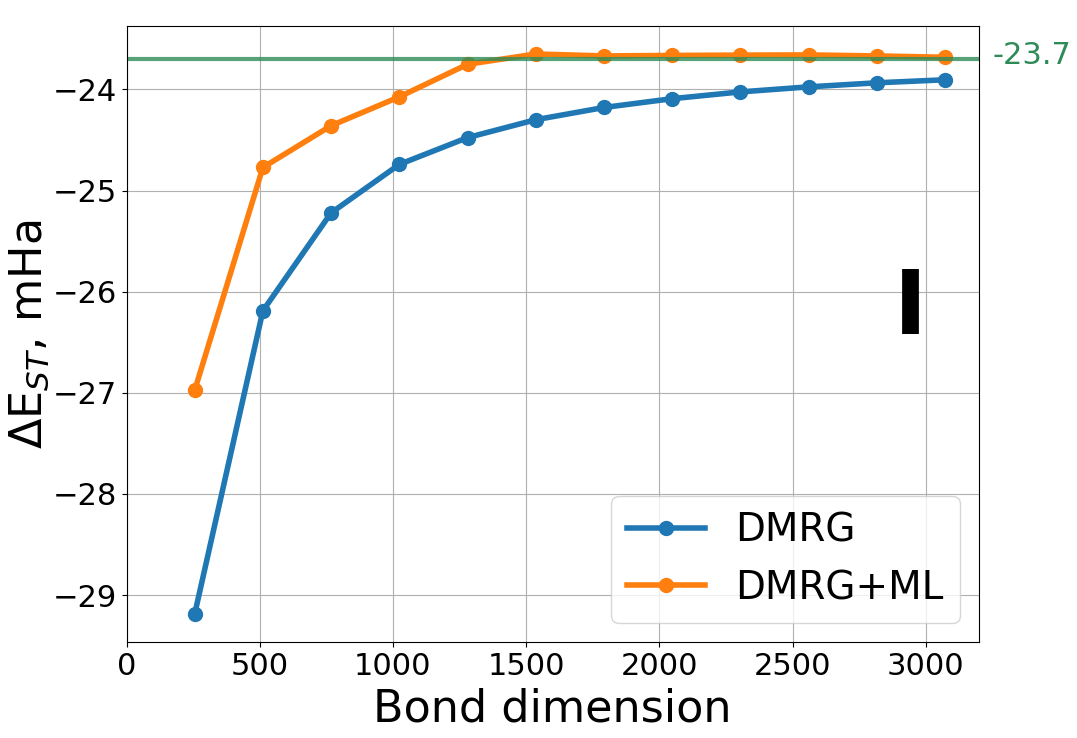}
\caption{Singlet-triplet (S-T) gap for [3]triangulene (\textbf{I}) computed at various bond dimensions. The reference S-T gap value of -2.375$\times$10$^{-2}$ Ha is obtained from a DMRG calculation with a bond dimension of 5000.}
\label{fig:S_triangulene_st_gap}
\centering
\end{figure}

\begin{figure}[h]
\includegraphics[width=12cm]{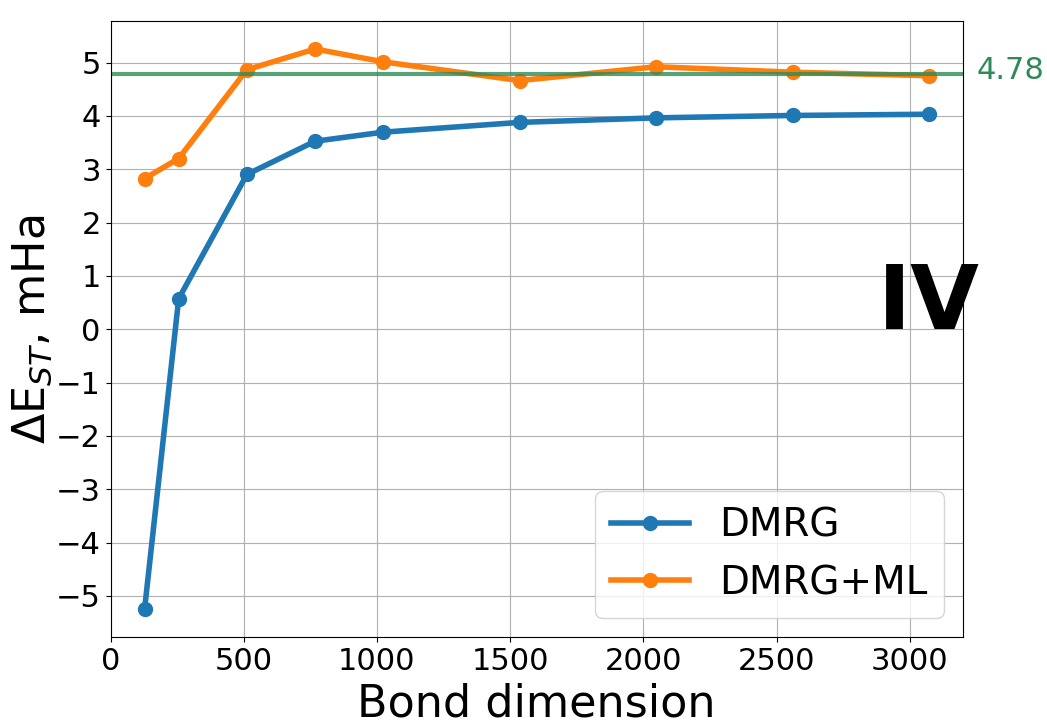}
\caption{Singlet-triplet (S-T) gap for peripentacene (\textbf{IV}) computed at various bond dimensions. The reference S-T gap value of 4.78$\times$10$^{-3}$ Ha is taken from S\'{a}nchez-Grande et al\cite{Grande-Urgel-2020}.}
\label{fig:S_perip_s1_t1_gap}
\centering
\end{figure}

\textbf{Model2} (Figure \ref{fig:test_mols_en_diff}--\ref{fig:S_perip_s1_t1_gap}) exhibits trends similar to \textbf{model1} discussed in the main text. At bond dimensions ranging from 750 to 1500, the difference between ML corrected energies and $E_{hb}$ energies falls below 1 mHa. The ML correction achieves an accuracy of 1 mHa in the worst case at truncation error of 5$\times$10$^{-5}$ (system \textbf{II}). However, for the other two examples, this accuracy is achieved already at truncation errors around 1$\times$10$^{-4}$ (Figure \ref{fig:test_mols_en_diff},(e)).

$\Delta$-ML DMRG procedure accurately predicts the S-T gap value for [3]triangulene at $M=1000$, Figure \ref{fig:S_triangulene_st_gap}. For peripentacene (Figure \ref{fig:S_perip_s1_t1_gap}), the S-T gap is correctly  predicted at $M=500$ with a truncation error of 1$\times$10$^{-4}$. Across bond dimensions from 500 to 3000, $\Delta$-ML DMRG approach provides stable predictions in the range of 4.8--5.1$\times$10$^{-3}$ Ha.

Final energies and truncation errors of all DMRG calculations performed on systems in training as well as test sets can be found in Ref. \cite{github}.

\bibliography{references}